
\magnification\magstep1
\baselineskip=16pt
\language=0
\def\no{\noindent}
\def\L{\Lambda}
\def\D{\Delta}
\def\s{\sigma}
\def\P{\Psi}
\def\bP{{\bar\Psi}}
\def\a{\alpha}
\def\ta{{\overrightarrow\tau}}
\def\b{\beta}
\def\bt{{\bar t}}

\def\en{\eqno}
\def\tmu{{\tilde\mu}}
\centerline{\bf Stability of Insulating Phases in the Hubbard Model:}
\centerline{\bf a Cluster Expansion}\par
\vskip 0.2in
\centerline{K. Ziegler}
\vskip 0.3in
\centerline{Institut f\"ur Theorie der Kondensierten Materie,
Universit\"at Karlsruhe,}
\centerline{ Physikhochhaus, D-76128 Karlsruhe, Germany}
\bigskip
\centerline{(\today)}
\vskip 0.6in
\noindent Abstract:\par
\no
The stability of the insulating regime of the Hubbard model on a
$d$-dimensional lattice, which is characterized by an
exponential decay of the Green's functions, is investigated in terms
of a cluster expansion. This expansion for the Green's function is organized
in terms of connected clustered transfer matrices. An upper bound for the
expansion terms is
derived for the hopping rate ${\bar t}$ depending on the coupling constant
$U$ as ${\bar t}<U/4d$. This implies an upper bound for the decay length of
the Green's function.
\bigskip
\no
\bigskip
\bigskip
\no
{\bf 1. Introduction}
\bigskip
\no
The Hubbard model was originally constructed to describe in a simple
way a metal-insulator transition for spin-dependent fermions (i.e.
electrons) [1]. This transition reflects the competition between
potential (static) energy and kinetic energy. The model is defined
on a lattice, where the potential energy consists of a chemical potential
and an on-site repulsion of fermions with opposite spin. The kinetic
energy is given by a nearest neighbor hopping. It turned out from
a number of calculations that this model has a rich structure because
of the complicated interplay of charge and spin degrees of freedom.
For instance, mean-field calculations for a magnetic order parameter
indicate para-, ferro- and antiferromagnetic states for
the half-filled system [2]. Thus, the magnetic properties of the model
became a central subject of investigations. The metal-insulator transition
was discussed originally by Hubbard using self-consistent approximations
[1], later in terms of a variational approach [3],
and more recently in the limit of an infinite dimensional lattice
[4]. However, the detailed mechanism and the properties of the
transition are not entirely clear at least for dimensionality $1<d<\infty$
[see, e.g., 2]. This problem is
partly related to the fact that the properties of the insulating state
are not completely understood.

To study the metal-insulator transition
one can, in principle, start either from the metallic or from the
insulating side. As the simplest approximations we could use non-interacting
fermions on the metallic side or the local (atomic) limit on the
insulating side, where the hopping rate is zero. Unfortunately, neither
of these starting points is very useful in order to understand the
interacting Hubbard model: Non-interacting fermions are unstable against
an arbitrarily weak interaction [2], and the local limit is
completely degenerated with respect to the spin. Therefore, an arbitrarily
weak hopping rate would lift the degeneracy leading to a new state
which might be magnetically ordered [5]. The basic idea of the present
work is to start from the insulating side without specifying its magnetic
order. Of course, this is not possible in terms of a mean-field theory
or by introducing a specific variational wave function. But we can
expand physical quantities of the Hubbard model, e.g. the free energy
or a Green's function, around the atomic limit (hopping expansion)
and estimate the asymptotic behavior on large scales. This is relatively
easy for fermions because the Pauli principle restricts the
number of expansion terms significantly. The expansion can be organized
in clusters of hopping processes and chains of sites which are empty,
singly or doubly occupied. The estimation procedure does not require
a knowledge of the magnetic order. Thus, we avoid the determination
of informations which might be difficult to obtain. On the other hand,
an insulating state is characterized by an exponential decay of the Green's
function which is related to the existence of a gap.
The exponential decay is a property which can be
obtain directly from the estimation of the hopping expansion. Therefore,
the stability of the insulating property can be studied with this method.
However, we cannot determine magnetic states and phase transition
inside the insulating regime.

The article is organized as follows:
In Sect.2 the Hubbard model is defined in a coherent state representation
for a grand canonical ensemble of fermions.
The partition function and the density of particles are calculated
in the local limit. Using this result a rough estimate for the probability of
a hopping process is given for a single particle. This simple argument leads to
a first estimate for the region of localization. At the end of Sect.2
the result of the estimation procedure of the hopping expansion for the
Green's function is summarized. Then in Sect.3 we present details of the
construction of the hopping expansion which appears as a cluster
expansion.  This includes the derivation of a
``Linked Cluster Theorem'' for the cluster expansion. It is different
from the usual Linked Cluster Theorem [7] because of the degeneracy of the
half-filled Hubbard model with respect to the spin degree of freedom.
Finally, we present the cluster expansion in a transfer matrix representation.
This makes the estimation of the expansion terms more transparent. In
particular, we eliminate the spin degree of freedom in the
estimation procedure (Sect.4). This leads eventually to an upper bound for the
hopping rate below which the hopping expansion is absolutely convergent. The
latter means that we are in a localized region of the Hubbard model.
\bigskip
\bigskip
\no
{\bf 2. The Hubbard Model}
\bigskip
\no
The Hamiltonian of the Hubbard model on a $d$-dimensional lattice
$\Lambda$ reads [2]
$$H[c_\a^\dagger(r),c_\a(r)]
=-\bt\sum_{<r,r'>}[c_\a^\dagger(r)c_\a(r')+h.c.]+\sum_r\Big[\mu
c_\a^\dagger(r)c_\a(r)-{U\over6}\Big(c^\dagger_\a(r)\ta_{\a,\b} c_\b(r)
\Big)^2\Big],\en(1)$$
where $c_\a^\dagger(r)$, $c_\a(r)$ are fermion creation and annihilation
operators, respectively. $\ta=(\tau_x,\tau_y,\tau_z)$ are Pauli matrices.
We have used the convention that repeated spin indices are summed
$\a=\pm1$, and $<r,r'>$ means nearest neighbor sites on $\Lambda$. $\mu$
is the chemical potential.

\no
Introducing the grand canonical ensemble of fermions,
the partition function of the Hubbard model reads in coherent state
representation [6]
$$Z=\int\exp(-S){\cal D}[\P,\bP]\en(2a)$$
with the action
$$S=i\D\sum_{r,t}{1\over i\D}\bP_\a(r,t)[\P_\a(r,t)-\P_\a(r,t-\D)]+
i\D\sum_t {1\over\hbar}H[\bP_\a(r,t),\P_\a(r,t-\D)].\en(2b)$$
The lattice points $r$ will be denoted as {\it sites}
in contrast to the points $(r,t)$ of the space-time lattice which will be
called {\it points}.
The Grassmann variables $\bP_\a(r,t)$ and $\P_\a(r,t)$ are independent.
Therefore, we can shift the time of these variables as
$$\P_\a(r,t)\to\P_\a(r,t+\D)$$
$$\bP_\a(r,t)\to\bP_\a(r,t)\en(3)$$
which implies a new action of the form
$$S=i\D\sum_{r,t}{1\over i\D}\bP_\a(r,t)[\P_\a(r,t+\D)-\P_\a(r,t)]+
i\D\sum_t{1\over\hbar}H[\bP_\a(r,t),\P_\a(r,t)].\en(4)$$
Therefore, the partition function of the Hubbard model reads with $\hbar=1$
$$Z=\int\exp\Big\{ -i\D\sum_{r,t}[{1\over i\D}
\bP_\a(r,t)\P_\a(r,t+\D)+(\mu-1/i\D)\bP_\a(r,t)\P_\a(r,t)]$$
$$+i\D\bt\sum_t\sum_{<r,r'>}\bP_\a(r,t)\P_\a(r',t)+i\D{U\over6}\sum_{r,t}
\Big(\bP_\a(r,t)\ta_{\a,\b}\P_\b(r,t)\Big)^2\Big\}{\cal D}[\P,\bP]\en(5)$$
with antiperiodic boundary conditions for the Grassmann field in
time-direction. The integration w.r.t. the product measure
for the Grassmann field $\P$, $\bP$
can also be written by implementing the local terms of the action as
$$d\mu(r,t)=\exp\Big\{-(i\D(\mu-1)\bP_\a(r,t)\P_\a(r,t)$$
$$+i\D{U\over6}\Big(\bP_\a(r,t)\ta_{\a\b}\P_\b(r,t)\Big)^2
\Big\}\prod_{\a=1}^2 d\P_\a(r,t)d\bP_\a(r,t).\en(6)$$
Thus, the partition function reads
$$Z=\int\exp\Big\{i\D\sum_{t=0}^\beta\Big[-\sum_r
{1\over i\D}\bP_\a(r,t)\P_\a(r,t+\D)+\bt
\sum_{<r,r'>}\bP_\a(r,t)\P_\a(r',t)\Big]\Big\}\prod_{r,t}d\mu(r,t).\en(7)$$
Expectation values, like the Green's functions, can be expressed in terms of
this lattice functional integral
$$\langle ...\rangle={1\over Z}\int ...\exp\{-S_{nl}\}\prod d\mu(r,t)
\en(8)$$
with the non-local part
$$S_{nl}=i\D\sum_{t=0}^\beta\Big[\sum_r{1\over i\D}\bP_\a(r,t)\P_\a(r,t+\D)
-\bt\sum_{<r,r'>}\bP_\a(r,t)\P_\a(r',t)\Big].\en(9)$$
For instance, the one-particle Green's function reads
$$G_{\s}((r',t'),(r'',t''))$$
$$=\langle\P_\s(r',t')\bP_\s(r'',t'')\rangle={1\over Z}\int \P_\s(r',t')
\bP_\s(r'',t'')\exp\{-S_{nl}\}\prod d\mu(r,t).\en(10)$$
\bigskip
\no
Neglecting the hopping of the fermions, the partition function separates
into a product of independent local terms as
$$Z=\int\prod_{r,t}\exp[-\bP_\a(r,t)\P_\a(r,t+\D)]
d\mu(r,t)$$
$$=\prod_r\int \prod_t\exp[-\bP_\a(r,t)\P_\a(r,t+\D)]
d\mu(r,t)=Z_1^N,\en (11)$$
where $N$ is the number of lattice sites.
$Z_1$ is the partition function of the Hubbard model with one lattice
site (atomic limit). The Berezin-integration of the Grassmann variables
of $Z_1$ yields
$$Z_1=\int\prod_t \exp[-\bP_\a(r,t)\P_\a(r,t+\D)]
d\mu(r,t)=1+2(1-i\D\mu)^{\beta/\D}+[(1-i\D\mu)^2-i\D U]^{\beta/\D}\en (12)$$
which leads in the continuous time limit ($\D\sim0$) to
$$\sim1+2e^{-i\beta\mu}+e^{-i\beta(2\mu+U)}.\en (13)$$
The factor 2 in front of the second term reflects the fact that the single
site partition function is degenerated with respect to the spin $\a=\pm1$.

The temperature formalism (which is obtained by the Wick rotation
$i\beta\to\beta$) allows for the evaluation of the fermion density as
$$n=2-{1\over \beta}{\partial\over \partial\mu}\log Z_1=2-2{1+e^{-\beta
(\mu+U)}\over2+e^{\beta\mu}+e^{-\beta(\mu+U)}}.\en (14)$$
This expression of $n$ is a consequence of the fact that the differentation
w.r.t. $\mu$ probes how many points are {\it not} occupied by a fermion.
(I.e., it gives zero if the site is occupied.)
In the zero temperature limit the density takes on three different values:
$$n\sim\cases{
0&$\mu<-U$\cr
1&$-U<\mu<0$\cr
2&$\mu>0$\cr
}.\en (15)$$
The jumps of the density are a consequence of the behavior of the energy
$E(\mu)$
$$E(\mu)={\partial\over \partial\beta}\log Z_1\sim\cases{
2\mu+U&$\mu<-U$\cr
\mu&$-U<\mu<0$\cr
0&$\mu>0$\cr
}\en (16)$$
which implies that $\partial E/\partial\mu$ is discontinuous.
Suppose that the chemical potential $\mu$ is restricted to $-U<\mu<0$
which means single occupancy of the site. Then it costs $\D E=\mu+U$ to add
another particle. Or it costs $\D E=-\mu$ to remove the particle from this
site.
Switching on the kinetic (hopping) term in the Hubbard model, the added or
removed particle can gain kinetic energy proportional to $\bt$ by undergoing a
hopping process. The probability of a hopping process consisting of $l$
successive hops is proportional to $(\bt/\D E)^l$. The latter means that the
hopping processes are irrelevant if $\bt<\min\{-\mu,U+\mu\}$; i.e., the
particles are localized in this case. This argument is not very strong
because it is based on single particle hopping processes. In the Hubbard
model, however, there is particle conservation: particles are always
accompanied by holes which hop with a different probability.
The idea of this article is to take the particle conservation
correctly into account in a rigorous estimation of the hopping expansion.
The result of this investigation, which will be given in detail below, can be
summarized in the following statement:
\bigskip
\no
{\it The Green's function $G_\s((r,t),(r',t))$ of the Hubbard model on a
$d$-dimensional lattice with chemical potential $\mu$ ($-U<\mu<0$) decays
exponentially for ${\bar t}<U/4d$ with a decay length
$\xi<-1/\log[4d{\bar t}/U]$.}
\bigskip
\no
This result means that the Hubbard model is in a localized phase for
sufficiently small hopping parameter. Thus, it gives a lower
estimate for the phase boundary of the insulating phase.
\bigskip
\bigskip
\no
{\bf 3. Cluster Expansion of the Hubbard Model}
\bigskip
\no
At first we will consider the expansion of the partition function of (7)
on a finite lattice $\L$ in terms of the space and time off-diagonal
contributions. In this case we do not experience convergency
problems, since we work with fermions which obey Pauli's principle.
The estimation of the expansion will be performed uniformly in the
size of the lattice such that the thermodynamic limit is included.
Considering the partition function in (7) or the expectation value
in (8) the hopping expansion can be developed from the Taylor expansion
of $\exp\{-S_{nl}\}$. The latter reads as a product w.r.t. nearest neighbor
bonds in space and time $<x,x'>$. Using the space-time notation $x=(t,r)$
we get
$$\prod_{<x,x'>}\exp[-\bP_\a(x)A(x,x')\P_\a(x')]$$
$$=\prod_{<x,x'>}\{1-
\bP_\a(x)A(x,x')\P_\a(x')+{1\over2}[\bP_\a(x)A(x,x')\P_\a(x')]^2\}.\en(17)$$
According to (9) we have
$$A((r,t),(r,t+\D))=1$$
and for $r,r'$ nearest neighbors
$$A((r,t),(r',t))=\D{\bar t}.$$
The expression on the r.h.s. of (17) can be expressed by an unperturbed part
$z_0(x,x')$ and a perturbation $z_1(x,x')$
$$\prod_{<x,x'>}z(x,x')=\prod_{<x,x'>}[z_0(x,x')+z_1(x,x')].\en (18)$$
The most interesting case for the hopping expansion is that where
the unperturbed system is degenerated w.r.t. the spins. This we find
if the lattice is singly occupied. The following discussion will
be restricted to this situation which appears for $-U<\mu<0$.
We choose for the unperturbed part singly occupied sites without hopping
$$z_0((r,t),(r',t))=1,\en (19a)$$
where $r,r'$ are nearest neighbors, and
$$z_0((r,t),(r,t+\D))=\bP_\a(r,t)\P_\a(r,t+\D).\en (19b)$$
I.e., the expansion of $z(x,x')$ around $z_0(x,x')$
at each bond leads to a cluster expansion for $Z$ or the Green's functions.
Thus, the set of bonds of the space-time lattice separates into a set of
clusters ${\cal C}$ where
the bonds are occupied by $z_1(x,x')$ and the complement ${\cal C}'$
where the bonds are occupied by $z_0(x,x')$. Such a contribution to the
expansion reads
$$\int\prod_{<x,x'>\in{\cal C}}z_1(x,x')\prod_{<x,x'>\in{\cal C}'}z_0(x,x')
\prod_xd\mu(x).\en (20)$$
Due to (19a,b) the unperturbed part of the partition function is
$$Z_0=\int\prod_{<x,x'>}z_0(x,x')\prod_xd\mu(x).\en (21)$$
In the cluster expansion we can now perform the integration over the
Grassmann variables $\P$, $\bP$ at each point $x$. There are three
different situations depending on the occupation of the points:

\no
empty point
$$w_0=\int d\mu(r,t)=(1-\D\mu)^2-\D U\sim e^{-2\D\mu-\D U}\en (22a)$$
singly occupied point
$$w_1=\int\bP_\a(r,t)\P_\a(r,t)d\mu(r,t)=1-\D\mu\sim e^{-\D\mu}\en (22b)$$
and doubly occupied point
$$w_2=\int\bP_1(r,t)\P_1(r,t)\bP_2(r,t)\P_2(r,t)
d\mu(r,t)=1\en (22c)$$
with the asymptotic behavior for $\D\sim0$.
\bigskip
\no
Thus, the statistical weights distinguish four regimes:

\no
(i) $\mu<-U$:
$$w_0>w_1>w_2$$

\no
(ii) $-U<\mu<-U/2$:
$$w_1>w_0>w_2$$

\no
(iii) $-U/2<\mu<0$:
$$w_1>w_2>w_0$$

\no
(iv) $0<\mu$:
$$w_2>w_1>w_0$$
\bigskip
\no
This implies for the unperturbed partition function
$$Z_0=(2w_1^{\beta/\D})^N\en (23)$$
which presents chains of singly occupied bonds along the time direction.
Each bond has the weight $w_1$. In the expansion there are also chains of
empty points (bonds) (indicated in Fig.1 by dotted vertical lines) and doubly
occupied points (bonds) (indicated in Fig.1 by double vertical lines)
and hopping elements with single and double hops (indicated by horizontal
lines). The restriction to these chains reflects Pauli's principle.
All possible situations at a single point $(r,t)$ are shown in Fig.1.
Combinations of these individual elements lead to space-time clusters of
empty and doubly occupied vertical lines and chains of hopping processes
obtained from the $z_1(x,x')$.
The partition function decays into products of clusters which are separated in
space. However, in time
they are not completely separated because the singly occupied lines carry
a spin degree of freedom. This is shown for the expansion around the singly
occupied system in Fig.2. The situation is more complicated compared
to the expansion around the empty system ($\mu<-U$) and the doubly
occupied system ($\mu>0$) where the coupling of clusters in the time
direction by a spin degree of freedom does not occur.
As in the case of classical systems or for the
perturbation theory around the empty quantum system there are several
ways to prove the linked cluster theorem. Here we will use the replica
trick, a method discussed in [6]. This method is based on the idea
that the logarithm of $Z$ can be expressed by the $n^{th}$ power of $Z$ as
$$\log Z={d Z^n\over dn}|_{n=0}.\en (24)$$
The formal introduction of the replicated model comes from a replication
of the Grassmann variables
$$\P_\a(x), \bP_\a(x)\to\P_\a^j(x), \bP_\a^j(x){\rm\ \ \ with\ \ }
j=1,2,...,n.\en (25)$$
This implies for (17)
$$\prod_{<x,x'>}\prod_{j=1}^n\exp[-\bP_\a^j(x)A(x,x')\P_\a^j(x')]$$
$$=\prod_{<x,x'>}\prod_{j=1}^nz^j(x,x').\en (26)$$
The bond expression $z(x,x')$ can be written again as a sum of an unperturbed
part and a perturbation $\prod_jz^j(x,x')=z_0(x,x')+z_1(x,x')$. For the
unperturbed part we write
$$z_0((r,t),(r',t))=1,\en (27a)$$
where $r,r'$ are nearest neighbors, and
$$z_0((r,t),(r,t+\D))=\prod_{j=1}^n[\bP_\a^j(r,t)\P_\a^j(r,t+\D)/2^{\D/\beta}
w_1].\en (27b)$$
$z_0$ is here normalized such that
$$Z_0=\int\prod_{<x,x'>}z_0(x,x')\prod_xd\mu(x)=1.\en (28)$$
Now we will show that the expansion of the partition function of a connected
cluster ${\cal C}$ of bonds, occupied by $z_1$ and surrounded by bonds occupied
with $z_0$, vanishes for $n=0$ except for the leading order. Using
${\cal C}_s$ as the surrounding of the
cluster ${\cal C}$ this partition function reads
$$Z_{{\cal C}}\equiv\prod_{j=1}^n\int\{\prod_{<y,y'>\in{\cal C}_s}
\bP_\a^j(y)\P_\a^j(y')\}\prod_{<x,x'>\in{\cal C}}z(x,x')\prod_{x\in{\cal C}_i}
d\mu(x),\en (29)$$
where ${\cal C}_i$ are the points which are connected by the bonds of the
cluster
${\cal C}$. $y$, in contrast to $y'$, in the first product does not belong to
${\cal C}_i$ or vice versa. The factors of the first product can be rearranged
in order to
extract the Grassmann variables which are not integrated. This leads to
$$\pm\{\prod_j\prod_y\bP_\a^j(y)\prod_{y'}\P_\a^j(y')\}\prod_j\int
\prod_{y''}\bP_\a^j(y'')\prod_{y'''}\P_\a^j(y''')\prod_{<x,x'>\in{\cal C}}
z(x,x')\prod_{x\in{\cal C}_i}d\mu(x).\en (30)$$
The $\pm$ in front comes from the fact that the Grassmann variables are
anticommuting, and is $+$ for an even and $-$ for an odd number of permutations
in the rearrangement. $y$ and $y'$ are now the points of ${\cal C}_s$
which do not belong to ${\cal C}_i$ whereas $y''$ and $y'''$ belong to
${\cal C}_i$. The product of the integrals gives $(Z_{{\cal C}})^n$.
The leading term of the cluster expansion inside the cluster ${\cal C}$
yields in leading order $(Z_{{\cal C},0})^n$ which is obtained from (30) by
replacing $z(x,x')$ by $z_0(x,x')$. For $n=0$ both expression agree and are
1. Thus, only the leading term of the connected cluster survives.

It is convenient to study the cluster expansion after an integration over the
Grassmann variables. The integration leads to a transfer matrix
representation of $Z$ or the Green's function. The transfer matrix connects
discrete hyperplanes at neighboring times. Thus, it describes hopping
processes as well as resting particles:
$$T(\{\nu_r\}|\{\nu'_r\})\en (31)$$
is a $4^N\times4^N$ matrix. $\nu_r=0,1,2,3$ denotes the
four possible states of a point: empty, singly occupied with two different
spin orientations and doubly occupied, respectively. The partition function
then reads
$$Z=Tr(T^{\beta/\D}).\en (32)$$
In particular, the transfer matrix of the unperturbed system is
$$T_0(\{\nu_r\}|\{\nu'_r\})=
\prod_r\delta_{\nu_r,\nu'_r}{1\over2}(\delta_{\nu_r,1}+\delta_{\nu_r,2})
\en (33)$$
which implies
$$Z_0=Tr(T_0^{\beta/\D})=1\en (34)$$
for the partition function.
The cluster expansion reads now in terms of the transfer matrix
$$T=T_0+T_1+T_2+...$$
where $T_1$,... are the contributions with empty or doubly occupied time-like
elements or hopping elements. The expansion terms of
$$Z=Tr[(T_0+T_1+...)^{\beta/\D}]\en (35)$$
read
$$Tr(T_{j_1}T_0^{k_1}...T_{j_m}T_0^{k_m}){\ \ \ \ }with{\ \ }\sum_{i=1}^m
(1+k_i)=\beta/\D.\en (36)$$
The trace decays into different clusters which have disconnected
projections onto space (see Fig.2b).
If the cluster expansion of the partition function decays into products of
disconnected clusters the corresponding expansion of the free energy $\log Z$
is an expansion in terms of connected cluster. This is the famous linked
cluster theorem [7]. There is an obvious separation of
disconnected clusters with respect to space because the trace can be
written as a product:
$$Tr(T^{\beta/\D})=\prod_lTr[T({\cal C}_l)],\en (37)$$
where $T({\cal C})$ is the product of transfer matrices on a space-time
cluster. The clusters $\{{\cal C}_l\}$ have a disconnected space projection.
The separation of clusters in time, which are connected by a spin degree of
freedom (e.g. Fig.2a), is less obvious. However, since we
have to take the derivative with respect to $n$ and set $n=0$ according to
(28), only contributions of the expansion of $Z^n$ survive which consist of
only one space-time connected cluster. This is the linked cluster property
for the transfer matrix representation
$$\log Z=\sum_{\cal C}g({\cal C}) Tr[T({\cal C})].\en (38)$$
The coefficient $g({\cal C})$ is $\pm1$ depending on the
special cluster ${\cal C}$. The replica trick can also be used to express a
fermion expectation value. For instance, the
Green's function $G_{\s}(x,x')$ can be written using
$$\Gamma_\s^j (x,x')=\int\bP_\s^j(x)
\P_\s^j(x')e^{-S_n}\prod_{r,t,j'} d\mu^{j'}(r,t)$$
$$=Z^{n-1}\int\bP_\s^j(x)
\P_\s^j(x')e^{-S}\prod d\mu(r,t)$$
$$\overrightarrow{n\to0}{\ \ \ }{1\over Z}\int\bP_\s^j(x)
\P_\s^j(x')e^{-S}\prod d\mu(r,t)=G_\s(x,x')\en (39)$$
where the last equation is just the definition of the Green's function
according to (10). In the expansion only clusters survive
which connect $x$ and $x'$ because the Grassmann variables appear in $S_n$
only in pairs. The limit $n\to0$ eliminates all
contributions which have clusters disconnected from that one which contains
the connection between $x$ and $x'$ because only the leading order term
contribute.
\bigskip
\bigskip
\no
{\bf 4. Estimation of the Connected Cluster Expansion}
\bigskip
\no
${\cal C}_{x,x';\s}$ is a connected space-time cluster containing the points
$x$ and $x'$ where these points are occupied by fermions with spin $\s$.
These are the clusters which contribute to the Green's function
$G_\s(x,x')$. The connected cluster expansion of the Green's function then
reads
$$G_\s(x,x')=\sum_{{\cal C}_{x,x';\s}}Tr[T({\cal C}_{x,x';\s})].\en (40)$$
The modulus of the Green's function can be estimated by the triangle
inequality as
$$|G_\s(x,x')|\le\sum_{{\cal C}_{x,x';\s}}|Tr[T({\cal C}_{x,x';\s})]|,
\en (41)$$
and applying the triangle inequality to the trace we obtain
$$\le\sum_{{\cal C}_{x,x';\s}}\sum_{\{\nu_r=0,3\}}
|T(\{\nu_r\}|\{\nu_r\}|{\cal C}_{x,x';\s})|.\en (42)$$
The summation goes over $\nu_r=0,3$, since only empty and doubly occupied
points occur at the boundaries of the cluster.
Finally, from the r.h.s. we pull out the maximum of the diagonal matrix
elements of the cluster transfer matrix as
$$\le\sum_{{\cal C}_{x,x';\s}}2^{|{\cal C}_{x,x';\s}|_{space}}
\max_{\{\nu_r=0,3\}}|T(\{\nu_r\}|\{\nu_r\}|{\cal C}_{x,x';\s})|\en (43)$$
where $|{\cal C}|_{space}$ is the number of sites of the space projection
of the cluster ${\cal C}$.
The summation can be separated into configurations of hopping elements in
space and time translations of the individual hopping elements.
All terms on the r.h.s. of (43) are positive. Therefore, we can sum over
larger cluster configurations which contain those actually present in (43).
(The simple (but insufficient) case of independent particle-hole hopping
processes is discussed in the Appendix.) In the following we will take into
account that we have particle conservation in a finite cluster; i.e.,
particles and holes are created in pairs. There are always two hopping events
if we create a new particle-hole pair. Given a
configuration of hopping elements in space, the cluster matrix
$T({\cal C})$ consists of vertical elements of length $t_0$
with weight $(w_0/w_1)^{t_0}\sim e^{-\D\mu_0t_0}$ (connecting empty points),
$(w_2/w_1)^{t_0}\sim e^{-\D\mu_2t_0}$ (connecting doubly occupied points) and
hopping elements. Thus, the expression
$\max_{\{\nu_r=0,3\}}|T(\{\nu_r\}|\{\nu_r\}|{\cal C}_{x,x';\s})|$
can be written as a product of the weight of the hopping element $\D{\bar t}$
and the weight of particle-hole pair $v=w_0w_2/w_1^2$. For a cluster with
$N$ hopping events (e.g. Fig.3) the maximal weight can be used as an upper
bound
$$\max_{\{\nu_r=0,3\}}|T(\{\nu_r\}|\{\nu_r\}|{\cal C}_{x,x';\s})|\le
(\D{\bar t})^Nv^{t_1+...+t_{N-1}},\en(44)$$
where $\{ t_j\}$ are the times between successive hopping events.
Now we sum over these times taking the continuum time limit $\D\to0$. This
leads to integrations in time
$$\lim_{\D\to0}(\D{\bar t})^{N-1}\sum_{t_1,...,t_{N-1}\ge0}v^{t_1+...+t_{N-1}}
$$
$$=({\bar t}\int_0^\infty v^tdt)^{N-1}=(-{\bar t}/\log v)^{N-1}=[{\bar t}
/(\mu_0+\mu_2)]^{-N+1}.\en(45)$$
Because of $\mu_0=\mu+U$ and $\mu_2=-\mu$ for $-U<\mu<0$ this bound is
independent of the chemical potential $\mu$
$$({\bar t}/U)^N.\en(46)$$
Thus, the r.h.s. of (43) (without a general factor $\D U$ which is absorbed
in the Green's function) has an upper bound
$$\sum_{N\ge|x-x'|}\sum_{c_N}2^{|{\cal C}_{x,x';\s}|_{space}}
({\bar t}/U)^N\le\sum_{N\ge|x-x'|}2^N\sum_{c_N}({\bar t}/U)^N
.\en(47)$$
The summation starts with $N=|x-x'|$ because this is the smallest cluster
which contains the connection of $x$ and $x'$. Here we assumed for simplicity
that $x'$ and $x$ have distance zero in time. A distance in time would imply
that one of the $t_j$ in the summation of (45) starts with the corresponding
$|t-t'|$. The inequality in (47) is due to the fact that the space projection
of a cluster with $N$ hopping
events has at most $N$ sites. $\sum_{c_N}$ is the summation
over space configurations of $N$ hopping events which form a connected cluster.
This gives the entropy of the different cluster configurations.
The connected clusters consist of closed paths of particles and holes because
there is creation and annihilation of particles and holes at the boundaries
of the clusters. The space projection of these closed paths must be connected
(otherwise it would not be a connected cluster) such that one can walk through
the whole space projected cluster. According to the K\"onigsberg bridge
problem [7] one can walk through the space projected cluster
visiting each path once and only once. Therefore, the configurations of
the space projected clusters can be generated by random walks. Then there
are at most $(2d)^N$ configurations for $N$ hopping events with a fixed
starting site, since each hopping event has $2d$ different choices to occur on
the $d$-dimensional cubic space. Thus, the summation $\sum_{c_N}$
leads to an additional
factor $(2d)^N$ when we estimate the sum with respect to all connected
cluster with $N$ hopping events and a given ``center of mass''. For the
latter we can use one site of the Green's function, $x$ or $x'$. Thus, the
connected cluster expansion converges absolutely if
$$2\cdot 2d\cdot{\bar t}{1\over U}<1.\en (48)$$
The first factor 2 comes from the spin degeneracy (eq. (43)), the factor $2d$
from the possibilities of a particle to hop away from a given site.
This implies a boundary value for the hopping parameter ${\bar t}_c
=U/4d$ below which $|G(x,x')|$ is bounded by a geometric series
$$\sum_{N\ge|x-x'|}\Big(4d\cdot{{\bar t}\over U}\Big)^N.
\en (49)$$
The upper bound decays exponentially with $|x-x'|$ with decay length
$-1/\log[4d{\bar t}/U]$. The exponential decay of the Green's function
means that the system is in a localized state.
\bigskip
\bigskip
\no
{\bf 5. Conclusions}
\bigskip
\no
Starting from the hopping expansion of the Hubbard model
on a $d$-dimensional cubic lattice we analyzed the convergence of
the expansion. The expansion of a Green's function, organized in terms
of connected clusters, provided a bound for the hopping rate $\bt$
below which there is an exponential decay (Fig.4).
This is a criterion for
the stability of the insulating regime. Moreover, we obtained an upper
bound for the decay length. The main advantage of the estimation method
was that we did not need informations about complicated spin-dependent
properties, the main obstacle in approximative investigations of the
Hubbard model. The expansion and its estimation were given for a one-particle
Green's function
with equal time. However, it could be applied to any other Green's function
of the Hubbard model using a similar estimation method.

It would be interesting to establish an analogous estimation procedure
for the perturbation theory around the non-interacting fermions in
order to have a better understanding of the metallic regime of the
Hubbard model.

\no
Acknowledgement

\no
I would like to thank P. W\"olfle for interesting discussions.
\vfill
\eject
\no
{\bf References}

1 J. Hubbard, Phys.Rev.Lett.3, 77 (1959), Proc.Roy.Soc. A276, 238 (1963);

{\ \ \ }ibid. A277, 237 (1964); ibid. A281, 401 (1964)

2 E. Fradkin, \sl Field Theories of Condensed Matter Systems\rm

{\ \ \ }( Addison - Wesley, Redwood City, 1991)

3 W.F. Brinkman and T.M. Rice, Phys.Rev. B2, 4302 (1970)

4 D. Vollhardt, \sl Correlated Electron Systems\rm

{\ \ \ } ed. V.J. Emery (World Scientific, Singapore)

5 Y. Nagaoka, Phys.Rev. 147, 392 (1966)

6 J. W. Negele and H. Orland, \sl Quantum Many - Particle Systems

{\ \ \ }\rm ( Addison - Wesley, New York, 1988)

7 J. Glimm and A. Jaffe, {\sl Quantum Physics} (Springer, 1987)

\vfill
\eject
\no
{\bf Appendix}

\no
As a simple case we choose cluster
configurations which are generated by independent sequences of empty and
doubly occupied lines for each site $r$. This corresponds to the situation
of independent hopping processes of holes and particles. We consider this
incorrect assumption in order to compare the corresponding hopping expansion
with the simple result at the end of Sect.2.

There are always two hopping
events if we change from empty to doubly occupied points and vice versa.
Only at the boundary of the clusters there is one hopping element because
it is changing by one particle from empty or double occupied site to
a singly occupied site.
On the other hand, the hopping elements are shared between neighboring
points. Therefore, each interval of points with the same occupation
can be associated with one hopping element $\D\bt$. Furthermore, we sum over
all possible sequences with a fixed number of hopping processes independently
for each site $r$. For a given set of sites this includes all cluster
configurations with the same space projection and a given number of hopping
events.
In the continuum time limit $\D\to0$ this summation becomes a time integral
$$\lim_{\D\to0}{\sum_{{\cal C}_{x,x';\s}}}^{(N)}
 \max_{\{\nu_r=0,3\}}|T(\{\nu_r\}|\{\nu_r\}|{\cal C}_{x,x';\s})|$$
$$\le{\bar t}^N\int_{t_0}^{t_1}e^{-\mu_2t}dt\int_{t_1}^{t_2}e^{-\mu_2t}
dt...\int_{t_{N-1}}^{t_{N}}e^{-\mu_{0,2}t}dt
\le\Big({{\bar t}\over\tmu}\Big)^N,\en (A.1)$$
where $\tmu=\min\{\mu_0,\mu_2\}$.
$\sum^{(N)}_{{\cal C}}$ is here the sum over all clusters with a given space
projection with $N$ hopping elements. The entropy of the different cluster
due to different hopping events yields analogously to Sect.4 the condition for
absolute convergence
$${\bar t}<\min\{-\mu/4d,(U+\mu)/4d\}=\cases{
(U+\mu)/4d&for $\mu<-U/2$\cr
-\mu/4d&for $\mu>-U/2$\cr
}.\en (A.2)$$
This is in agreement with the simple argument given at the end of Sect.2.

\vfill
\eject
\no
{\bf Figure Captions}
\bigskip
\no
Fig.1

\no
Elements of the hopping expansion at a given time space point.
Vertical direction is time and horizontal direction is space direction.
Dotted lines connect empty points, single lines singly occupied points and
double lines doubly occupied points. Horizontal lines denote hopping events.
\bigskip
\no
Fig.2

\no
Examples for clusters as a result of the hopping expansion. Fig.2a shows two
clusters which are connected in time at site $r$ only through the spin
carrying singly occupied line. Fig.2b shows the same two clusters which are
disconnected in time because they are shifted by one space unit relative to
each other.
\bigskip
\no
Fig.3

\no
Examples for clusters with the same number of hopping events. Cluster (a) has
a higher weight than (b) because $v<1$ (see text).
\bigskip
\no
Fig.4

\no
Localized regime in the phase diagram of the Hubbard model: hopping rate $J$
($={\bar t}/U$) versus chemical potential $m$ ($=\mu/U$). The
triangular region I is obtained from a weak estimate based on independent
hopping processes of particles and holes, whereas the rectangular region II
is the result of the estimation with particle conservation.

\bye